# Design of a High-Performance Tomographic Tactile Sensor by Manipulating the Detector Conductivity

Shunsuke Yoshimoto, Membership, Koji Sakamoto, Rina Takeda, and Akio Yamamoto, Membership

*Abstract*—Recent advancements in soft robots, human-machine interfaces, and wearable electronics have led to an increased demand for high-performance soft tactile sensors. Tomographic tactile sensor based on resistive coupling is a novel contact pressure imaging method that allows the use of an arbitrary conductive material in a detector. However, the influence of material properties on the sensing performance remains unclear and the efficient and appropriate selection of materials is difficult. In this study, the relationship between the conductivity distribution of the material used as a detector and the sensing performance including sensitivity, force range, spatial resolution, and position accuracy is clarified to develop a high-performance tomographic tactile sensor. The performance maps reveal that a material with a conductivity of approximately 0.2 S/m can serve as an effective detector for touch interactions involving a force range of several Newtons. Additionally, incorporating gradient conductivity in the cross-section of the detector and multi-layer conductive porous media with anisotropic conductive bonding can help expand the design flexibility for enhanced performance. Based on these findings, various tomographic tactile sensors for soft grippers, tangible input interfaces, flexible touch displays, and wearable electronics are demonstrated by using a conductive porous media.

## I. Introduction

HUMAN skin demonstrates excellent pressure detection sensitivity and a wide range of pressure detection capabilities that can be leveraged in various touch interactions. Tactile sensors that mimic these functionalities and abilities have emerged as promising tools for controlling robotic fingers and hands [1], [2], designing prosthesis [3], recognizing objects touched by humans [4], implementing touch input user interfaces [5], and developing haptic feedback systems [6]. To realize these functions, the sensitivity, force range, spatial resolution, and accuracy of the sensors must satisfy the application requirements. Additionally, tactile sensors for soft robotics must exhibit flexibility, scalability, and shape complexity.

Manuscript received Month xx, 2xxx; revised Month xx, xxxx; accepted Month x, xxxx. This study was partially supported by the Japan Society for the Promotion of Science KAKENHI Grant Numbers JP22H03626 and Nitto Denko Corporation.

(Authors' names and affiliation) S. Yoshimoto and A. Yamamoto are with the Graduate School of Frontier Sciences, The University of Tokyo, 5-1-5 Kashiwanoha, Kashiwa-shi, 277-8563, Chiba, JAPAN (e-mail: yoshimoto@k.u-tokyo.ac.jp).
R. Takeda and K. Sakamoto are with the Corporate R&D division, Nitto Denko Corporation, 1-1-2 Shimohozumi, Ibaraki-shi, 567-8680, Osaka JAPAN.

Conventional tactile sensing has been commonly realized using arrayed or matrix-type pressure-sensitive sensors based on capacitive or resistive principles [7]–[12]. Additionally, camera-based sensing [13], [14] has proven useful for accurate pressure imaging. The magnetic approach offers advantages in sensor shape flexibility and deformability [15], [16]. However, a systematic approach for designing and fabricating a tactile sensor based on the required performance is lacking. In particular, most design strategies do not simultaneously consider the sensitivity, force range, spatial resolution, and accuracy of tactile sensors.

Tomographic tactile sensors based on soft materials and the principle of electrical impedance tomography (EIT) [17] have demonstrated advantages in flexibility and scalability [18]–[21]. Such sensors enable the visualization of the impedance distribution associated with contact pressure by capturing potential data using multiple electrodes placed arbitrary on a pressure-sensitive material. Recently, an innovative framework, a tomographic tactile sensor using resistive coupling between two conductors has been proposed [22]. Notably, in such sensors, arbitrary conductive materials with various conductivities can be used instead of conventional pressure-sensitive materials, allowing the realization of varied performance metrics by using different conductive materials. However, the material properties required for high-performance sensing remain to be extensively explored.

For the design and application of effective tomographic tactile sensors, it is crucial to understand the relationship between the performance metrics, i.e., the sensitivity, detectable range, spatial resolution, and position accuracy, and controllable material properties, such as the conductivity and elasticity. In particular, by creating a performance map, we can select effective materials for tomographic tactile sensors based on the required performance characteristics.

Considering these aspects, this research aimed at leveraging the advantages of tomographic tactile sensors based on resistive coupling and clarifying the relationship between the material properties and sensor performance. Inspired by prior work related to material design for enhancing energy efficiency [23], [24], we implemented a conductivity gradient in the cross-section to extend performance range, given that the potential distribution within a detector material depends on its conductivity distribution. Additionally, given the recent trend of advanced soft sensors based on porous media [25]–[28], we introduced conductive porous media owing to its lightweight, flexible, deformable, and easy to process nature. Figure 1 illustrates the design and sample application of a tomographic tactile sensor based on resistive coupling. Our finding provides a strategy to select a material for soft tactile sensors, allows to implement various application in soft robotics.

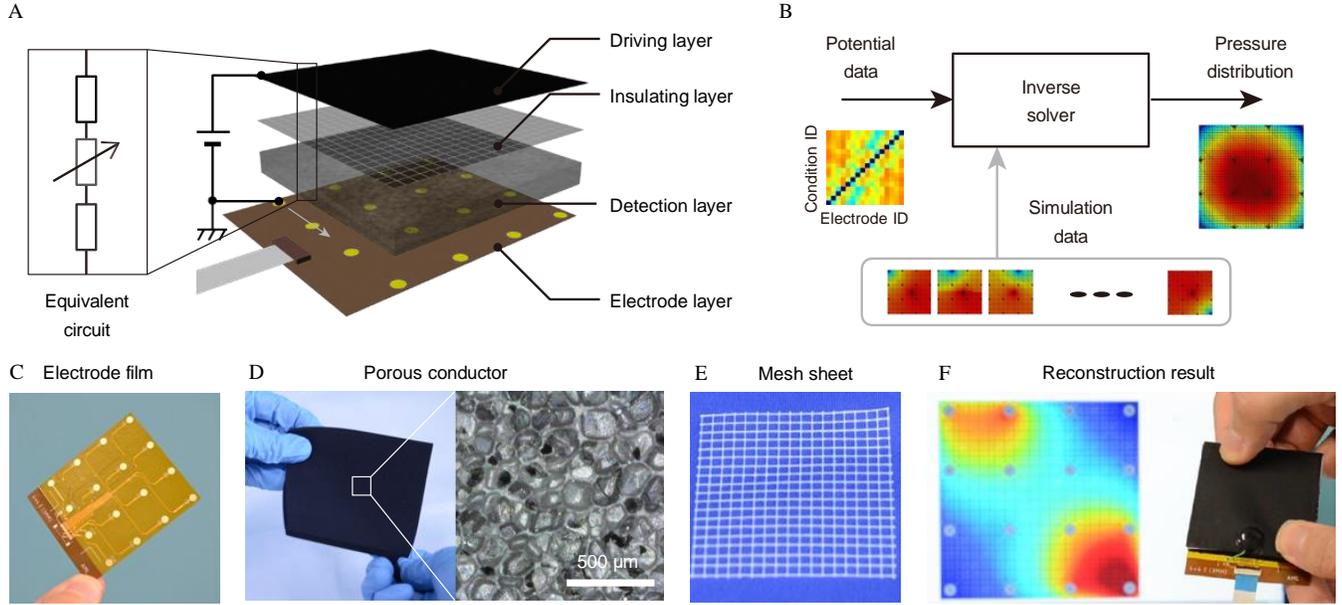

Fig. 1. Assembly and sensing framework of a tomographic tactile sensor based on resistive coupling. (A) Schematic of the detector and equivalent circuit. A DC voltage source is connected to the driving layer and the grounded electrode is sequentially switched. (B) Potential data, e.g., a 16 × 16-D potential vector, is processed using a reconstruction algorithm such as a linear regularization method or convolutional neural network (CNN) method. (C) Electrode film used for demonstrating the sensor. Sixteen electrodes with a diameter of 3 mm are located in a 4 × 4 array. (D) Conductive porous material (Sample A in Table I) used for the detection layer. Microscopic image captured using a digital microscope (Dino-Lite, Pro AD7013MT). (E) Mesh sheet (13-7194, KLASS Corp.) typically used to prepare insulating layers. (F) Reconstructed pressure distribution using a pressure imaging sensor sized 50 × 50 mm$^2$.

## II. Tomographic Tactile Sensor based on Resistive Coupling

### A. Overview

Figure 1(A) shows the structure of the tomographic tactile sensor based on resistive coupling. The detector of the sensor consists of multiple electrodes, a detection layer, an insulating layer, and a driving layer. Resistive coupling introduced between the detection and driving layers. Therefore, the detection and driving layers are conductive, whereas the insulating layer is a mesh structure or a sheet with multiple holes, inducing a small gap between the conductive layers. When the driving and detection layers are in contact, a potential distribution is produced on the detection layer based on the pressure distribution, captured by each reference electrode. To accurately reconstruct a pressure distribution from the potential data derived from sparse sampling points, sequential measurements are obtained under multiple grounding conditions. Typically, one electrode serves as the ground, selected using a multiplexer and cycled through all electrodes. Various methods can be used for the inverse solver, including linear regularization [29], block sparse Bayesian learning [30], and CNN method [31], [32].

### B. Analytical Model

Because the sensor relies on the resistive contact for detection, the sensitivity characteristics can be expressed using an analytical model. To formalize the input–output characteristics of the resistance-based tomographic sensor, we introduce an electromechanical model. According to the contact theory of the elastic plate and hemisphere, the relationship between the contact force $F$ and contact area $S_c$ can be expressed as follows:

$$S_c = \left( \frac{1-v^2}{E} \frac{3r}{4} \right)^{\frac{1}{3}} F^{\frac{1}{3}}, \quad (1)$$

where $r$ is the curvature of the contacting object, $v$ denotes Poisson's ratio, and $E$ is Young's modulus. The relationship can be generalized using a power function:

$$S_c = aP^\gamma. \quad (2)$$

The relationship between the contact area and contact resistance can be expressed as follows:

$$R_c = \frac{l}{\sigma S_c}, \quad (3)$$

where $\sigma$ and $l$ denote the conductivity and thickness of the detector surface, respectively. The voltage at the contact area can be expressed as follows:

$$\varphi = \frac{R_0}{R_c + R_0} V_{cc}, \quad (4)$$

where $R_0$ is the volume resistance of the detector. Finally, we can obtain the following relationship.

$$\varphi = \frac{R_0 a F^\gamma}{1 + R_0 a F^\gamma} V_{cc}. \quad (5)$$

We conclude that the sensitivity increase with a higher conductivity at the top of the detector and lower conductivity at the bottom of the detector. Moreover, reduced elasticity of the detector corresponds to a higher sensitivity.

### C. Inverse Solver

As shown in Figure 1(B), a tomographic tactile sensor requires a solver to determine pressure distribution from sparse potential vectors. To facilitate the finite element method simulation and solve the inverse problem, we implemented a software based on an existing method [22]. Specifically, we used Jacobian data calculated using a thin shell model with a uniform conductivity to reduce the calculation cost. Moreover, the use of a thin shell model eliminates the influence of conductivity of the model on the calulated Jacobian values. The effects of the use of a thin shell model for a thick detector are investigated in simulation section. The model consists of approximately 4000 triangles and 2000 nodes. The conductivity of the model was set as 1 S/m, and the driving voltage was set as DC 2 V. To gain fundamental insights, we used a linear reconstruction method, i.e., the Tikhonov regularization method [33]. Jacobian data for the reconstruction was obtained using the simulation, and the parameter $\lambda^2$ =5000 was selected for the simulation and experimental studies.

### D. Performance Metrics

In this paper, the sensing performance was evaluated using four metrics, i.e., sensitivity, maximum force, spatial resolution, and position accuracy, calculated using MATLAB based on the following definitions.

1) Sensitivity: Sensitivity indicates the rate of change of the sensor output at the centroid of the deconstructed image relative to the applied force. Because the sensor has a nonlinear output profile, we fit the output model to the data using the Levenberg–Marquardt method and calculate the derivative at the half value of the maximum applied force. The analytical model (5) can be simplified as follows:

$$\varphi(F) = \frac{p_1}{F^{p_3} + p_2}, \quad (6)$$

where $p_1$, $p_2$, and $p_3$ are constants. The sensitivity can be expressed as the differential value at the half value of the maximum force $F_h$.

$$SENS = \frac{d\varphi(F_h)}{dF} = -\frac{p_3 p_1 F_h^{p_3 - 1}}{(F^{p_3} + p_2)^2}. \quad (7)$$

2) Maximum force: The maximum force is the largest value of the detectable force. We estimate the force value at 90 % of the saturation output as follows:

$$FMAX = \varphi^{-1}\left(0.9\frac{p_1}{p_2}\right) = \left(\frac{p_2}{0.9}\right)^{\frac{1}{p_3}}. \quad (8)$$

3) Spatial resolution: The spatial resolution represents the spread width of the reconstructed image using a single-point input, calculated as the maximum distance between the center of the reconstructed potential input distribution and contour line indicating 50% of the maximum value of the reconstructed data. In the simulation and actual measurement, we used a contactor with a diameter of 5 mm, which is adequately smaller than the length of the detector and can be regarded as a single point input. The spatial resolution can be expressed in terms of the full width at half maximum ($FWHM$):

$$SR = 1 - FWHM/WIDTH, \quad (9)$$

where $WIDTH$ is the sensor width.

4) Position accuracy: The contact position $P_{cop}$ is defined by the centroid of the reconstructed image. The position accuracy is the distance between the true and estimated contact positions and expressed as

$$PA = 1 - (|P_{cop} - P_t|)/WIDTH, \quad (10)$$

where $P_t$ is the true position.

## III. Simulation Study

### A. Models and Conditions

A Simulation study was conducted to investigate the influence of the detector conductivity on the detection performance. The model of the tomographic tactile sensor is shown in Figure 2. We solved the Laplacian elliptic partial differential equation to identify the potential distribution on 60 mm square detector with a height of 10 mm. Sixteen electrodes with a diameter of 4 mm were arranged in a grid pattern on the bottom surface of the detector. For the driving layer, we applied a DC 2 V voltage to a single contact region with a diameter of 4 mm, located at the top of the detector. This simplification was aimed at eliminating the influence of the mechanical characteristics. To explore an effective configuration expanding the detection performance range, we introduced a gradient conductivity to the detection layer. The characteristics of this layer can be expressed as

$$\sigma(y) = \sigma_{low} \left(\frac{\sigma_{up}}{\sigma_{low}}\right)^{\frac{y - y_{min}}{y_{max} - y_{min}}}, \quad (11)$$

where $\sigma_{low}$ and $\sigma_{up}$ are the lower and upper conductivities, respectively; and $y_{min}$ and $y_{max}$ are the bottom and top vertical positions of the sensor, respectively.

The conductivity range was set from 0.001 to 100 S/m and divided into 11 equal intervals on a logarithmic scale. Therefore, the performance was calculated for 121 different conditions. Output voltage from the 16 electrodes was measured for each grounding condition, aiding the reconstruction of the potential distribution using the simulated data. To simulate various contact pressures, we varied the conductivity of the elements under the contact region from 0.001 to 10 S/m on a logarithmic scale.

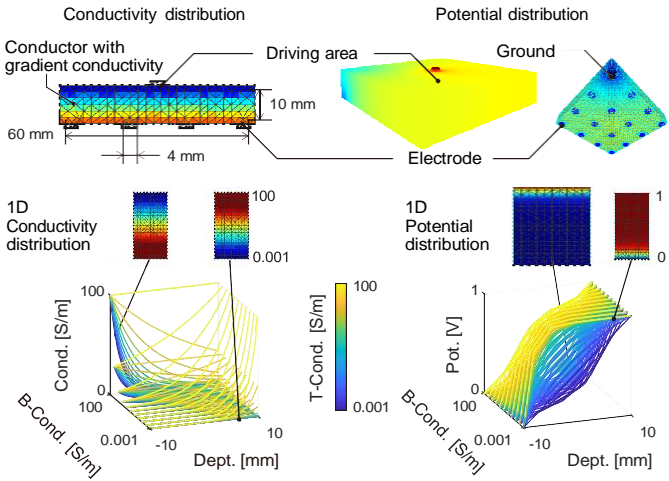

Fig. 2. Simulation model involving a detector sized 60 × 60 × 10 mm$^3$ with 16 electrodes at the bottom of the material and a single driving area at the top. The detector incorporates an exponent gradient distribution perpendicular to the plane, with 11 conductivity conditions for both the top surface (T-Cond.) and bottom surface (B-Cond.). The corresponding potential (Pot.) distribution in 1D is shown in the right plot.

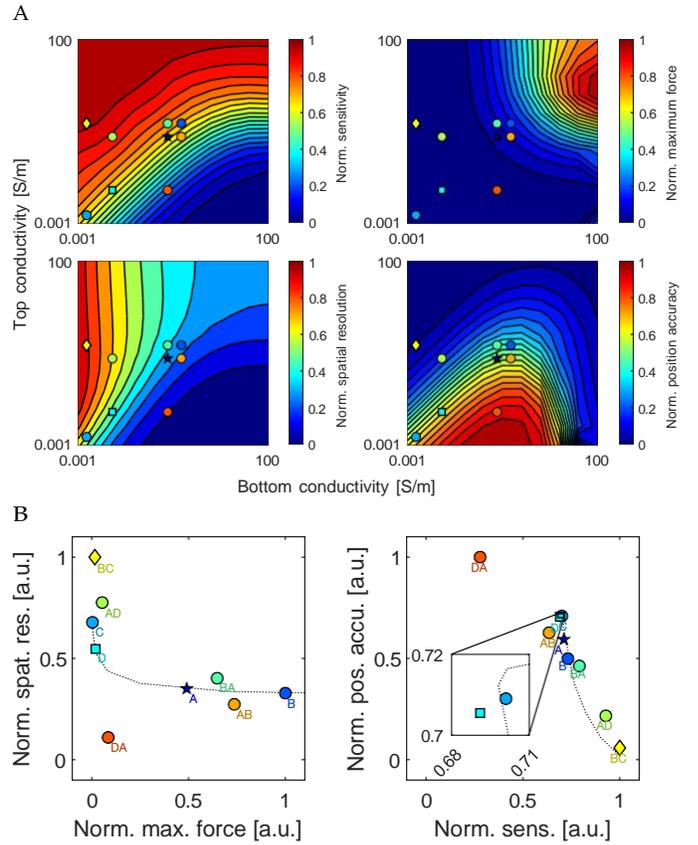

Fig. 3. Results of a simulation study evaluating sensor performance under various conductivity values of the detection layer. (A) Resulting contour maps of the evaluation metrics, i.e., the sensitivity (sens.), maximum force (max. force), spatial resolution (spat. res.), and position accuracy (pos. accu.) maps for selected samples. These metrics are normalized. (B) Positioning plots of selected materials (Table I) in the spatial resolution and maximum force space and in the position accuracy and sensitivity space. These metrics are normalized. The dashed line represents the performance curve based on uniform conductivity.

### B. Results and Discussion

To examine the influence of the detector conductivity on the sensing performance, four evaluation metrics, i.e., sensitivity, maximum force, spatial resolution, and position accuracy were determined for various conditions (detailed in performance metrics section). Figure 3(A) shows the normalized sensor performance based on a simulation study involving different conductivity values of the detection layer. Our findings highlight that a conductivity of approximately 0.2 S/m is suitable for detecting human touch interactions. Conversely, for ultra-high sensitivity sensors, a configuration with gradient conductivity of 0.5 S/m (top side) and 0.001 S/m (bottom side) is recommended. The sensitivity and maximum force were calculated by fitting curves mapping the relationship between the maximum potential and conductivity of the driving region. The sensitivity map indicates that the higher conductivity of the top surface corresponds to a higher sensitivity, consistent with the results of the analytical model. However, excessive conductivity leads to saturation of the force characteristics owing to the notable impact of the conductivity of the driving layer. Spatial resolution depends on the potential magnitude, with a lower conductivity at the bottom surface corresponding to a superior spatial resolution. A larger maximum detectable force can be achieved using a material with high and uniform conductivity compared with that with a gradient conductivity. To increase the position accuracy, the conductivity at the bottom side must be higher than that of the top side. Moreover, the ratio to the conductivity of the driver layer also plays a key role, with the conductivity of the bottom surface exhibiting a peak at approximately 1 S/m.

Figure 3(B) shows the positioning plots of the materials used in the experiment. Sample A demonstrates balanced performance, sample BC exhibits the highest sensitivity and spatial resolution, and Sample DA displays the highest position accuracy. The results highlight that there are no conditions that maximize all metrics simultaneously. In particular, a trade-off relationship exists between the sensitivity and maximum force, the position accuracy and maximum force, and the sensitivity and position accuracy. The use of a material with gradient conductivity helps expand the design flexibility for the performance enhancement. The simulation results are applicable for detectors with the different sizes of the detector because the potential distribution is not affected by the size. However, the ratio of the length to the thickness of the detector affects its performance as shown in Figure 4. The results indicate strong correlation between the potential vectors associated with a thin shell model and thick model. The error increases with the detector thickness.

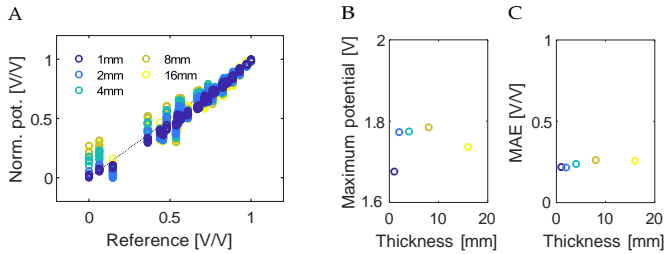

Fig. 4. Effect of the thickness of the detector (simulation study). Normalized potential vectors for various thicknesses: 1, 2, 4, 8, and 16 mm are calculated. The conductivities of the detector and driving area are set as 1 S/m and 1 S/m, respectively. (A) Correlation between the reference and potential vector of each thickness. As reference, the potential vector associated with a thin shell model is represented by a dashed line. (B) Relationship between the maximum potential and thickness. (C) Relationship between the mean absolute error (MAE) and thickness.

## IV. Measurement and Characterization

### A. Detector Fabrication

To explore the fundamental performance of the material, we fabricated the detector with a simple configuration. The detector consisted of a rigid electrode board and conductive porous media. The electrode board had 16 electrodes, as in the simulation configuration. Four porous materials with diverse conductivities were selected to validate the simulation results (Table I). A thompson blade (0.5 mmt) was used to cut the materials into samples with areas of 60 × 60 mm$^2$ and thickness of 5 mm. To incorporate conductivity gradient in the thickness direction, the materials were stacked. Notably, when two conductors are bonded together using surface adhesive, the conductivity is affected by the conductivity in the surface direction of the adhesive used. To overcome this problem, we used an anisotropic conductive bonding technique [34] based on conductive bonding of dot patterns (Figure 5(A)). To evaluate the effect of the dot patterns, we prepared two bonding conditions, i.e., 5 × 5 dots and 7 × 7 dots. The dot diameter was set as 7.22 mm. The conductive porous material was stacked and connected to the electrode by using a mask prepared by double-coated adhesive tape (No.5000NS, Nitto Denko Corp.) and conductive glue (CW2401, Circuit Works).

### B. Experimental Setup

The sensing circuit consists of a microcontroller (ESP-WROOM-32, Espressif Systems) and an analog switch (ADG726, Analog Devices). We used 16-channel 12-bit analog inputs for measuring the potential at each electrode. The analog switch was controlled using four-channel digital outputs from the microcontroller for selecting a grounding electrode from 16 electrodes. Because 16 electrodes were used, a single frame could capture 256 voltage data points (16 electrodes × 16 conditions). The sensing circuit was connected to the detector using a flat flexible cable. The circuit diagram and implemented circuit boards are shown in Figure 5(B). The loading equipment is illustrated in Figure 5(C). For the driving layer, a conductive rubber sheet (EC-20BH, Shin-Etsu Chemical Co., Ltd.) was attached to the contactor to maintain the repeatability of the loading condition. The contact position and force relative to the detector fixed on the plate could be precisely controlled using the apparatus, i.e., the XY stage (KYL06100-N2-F1-L, SURUGA SEIKI Co., Ltd.) and linear actuator (DRLM42G-04A2P-K, Oriental Motor) controlled using an AD/DA converter (USB-6216, National Instruments). A calibrated 3D force sensor (USL06-H5-50N-A, Tec Gihan Co.,Ltd) was installed between the actuator and indenter. The indenter with a diameter of 5 mm was moved using the actuator at a rate of 0.25 mm/s and applied force of 0 – 5 N. The nine contact positions, covering all combination of -20, 0, and 20 mm, were used for the experiment. The control and recording system was implemented using LabView (National Instruments).

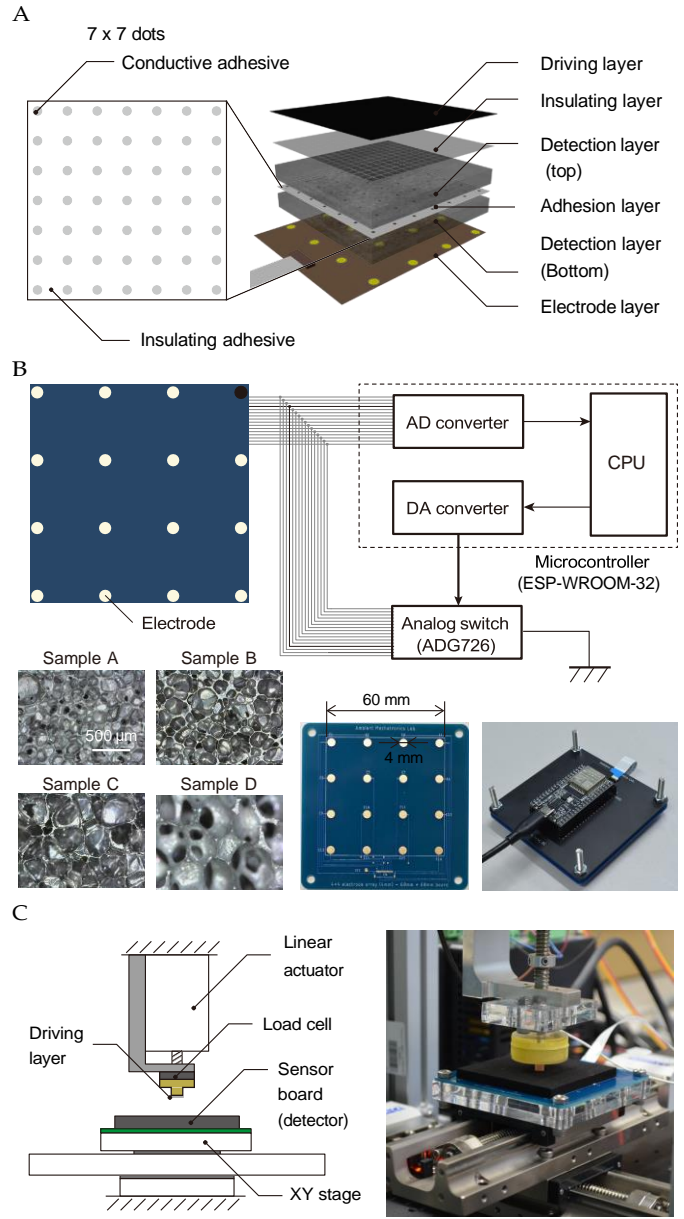

Fig. 5. Experimental setup. (A) Schematic of the sensor structure using a stacked detection layer. Two conductive materials stacked using the conductive adhesive pattern. (B) Circuit diagram, materials and boards used in the actual measurement. The electrode board has 16 electrodes with a diameter of 4 mm. The circuit board consists of a microcontroller and an analog switch. Microscopic images captured using a digital microscope (Dino-Lite, Pro AD7013MT). (C) Overview of the experimental setup. The detector of the proposed sensor was positioned on the XY stage. A contactor attached to the linear actuator applied force to the surface of the detection layer.

TABLE I
Electromechanical properties of samples used in the experiment. The surface resistance was measured using a low-resistance resistivity meter (Loresta-GX(Ⅱ) MCP-T710, Nittoseiko Analytech Co., Ltd.). The elasticity was measured using a universal testing machine (Autograph AGX-V2, Shimadzu Corporation).

| Sample | A | B | C | D |
|---|---|---|---|---|
| Type | C-4255E | E4385 | TK-P2N2 | F-10 |
| Manufacturer | INOAC | INOAC | SANWA SUPPLY | HOZAN |
| Main ingredient | Conductive CR (polychloroprene) | Conductive EPDM (ethylene propylene diene monomer) | Conductive PU (polyurethane) | Conductive PU (polyurethane) |
| Surface resistance [Ω/sq] | 880 | 384 | 120000 | 25000 |
| Conductivity [S/m] | 0.2273 | 0.5208 | 0.001667 | 0.008 |
| Elasticity [MPa] | 0.35 | 0.96 | 1.51 | 2.23 |
| Shore Hardness | 20 | 30 | 30 | 30 |

TABLE II
Comparison of single-layer and single-layer detectors. Force – output profiles show the relationship between the reconstructed value (out.) at the centroid and force at the center of the detector. The data are divided into the loading and unloading states, represented by circles and triangles with faint colors, respectively. Line profiles show the normalized reconstructed values (norm. out.) for the loading state on the line through the centroid. In the plots of the estimated position, the cross mark represents the contact location, i.e., ground truth, whereas the circle represents the centroid at the loading force of 2.25 N. The error vector for each contact location is presented. The position is normalized within the range of -1 to 1.

| Single-layer sample | A | B | C | D |
|---|---|---|---|---|
| Sensitivity [mV/N] | 0.174 | 0.011 | 0.108 | 0.102 |
| Maximum force [N] | 2.837 | 6.008 | 4.356 | 34975 |
| Spatial resolution [mm/mm] | 0.711 | 0.394 | 0.417 | 0.788 |
| Position accuracy [mm/mm] | 0.697 | 0.745 | 0.783 | 0.562 |
| Force – output | | | | |
| Line profile | | | | |
| Estimated position | | | | |

| Multi-layer sample | BA | AD | BC | AB | DA | BB - 5 × 5 | BB - 7 × 7 |
|---|---|---|---|---|---|---|---|
| Sensitivity [mV/N] | 0.009 | 0.223 | 0.418 | 0.017 | 0.064 | 0.009 | 0.005 |
| Maximum force [N] | 5.206 | 2.228 | 1.392 | 13.71 | 105.5 | 7.032 | 8.525 |
| Spatial resolution [mm/mm] | 0.522 | 0.528 | 0.442 | 0.454 | 0.736 | 0.381 | 0.466 |
| Position accuracy [mm/mm] | 0.752 | 0.620 | 0.625 | 0.607 | 0.626 | 0.732 | 0.804 |
| Force – output | | | | | | | |
| Line profile | | | | | | | |
| Estimated position | | | | | | | |

## C. Results and Discussion

*1) Effect of the detector conductivity on sensing performance:* The experimental results (measurement and characterization section) of the four samples and five double-layer samples are summarized in Table II. The positioning plots of the samples using the normalized value of each metric are shown in Figure 6. The results of certain samples, i.e., A, DA, and BC, match the simulation trends. The inconsistent trends of the other samples may be attributable to the factors not considered in simulations, e.g., the microscopic surface texture, uniformity of conductivity, uniformity of adhesion, and differences in elastic modulus. Specifically, the performance of sample D appears to be affected by the presence of holes (Figure 5(B)) and higher elasticity. To bridge the gap between the simulation and actual measurement results, non-porous media, such as a conductive silicone, can be used.

*2) Fabrication of porous material with gradient conductivity:* According to the examination of the influence of the detector conductivity on the sensing performance by the simulation study, material with gradient conductivity along its cross-section expands the performance design flexibility of the tomographic tactile sensor, as shown in Figure 3(B). For example, the highest sensitivity is achieved by stacking Samples B and C, and a balanced performance between Samples A and B is attained by stacking Samples A and B. Although several methods have been proposed to fabricate a material having gradient conductivity [35]–[37], it is challenging to maintain uniform conductivity in the horizontal direction. Therefore, we established an alternative method of fabricating a material with gradient conductivity by stacking uniform materials. To further clarify the characteristics of the anisotropic conductive bonding, simulation and experimental studies were performed.

Figure 7(A) shows the relationship between the position error and dot diameter of the adhesive, obtained through simulation. Larger diameters and larger numbers of dots correspond to a smaller position error. The local peaks observed for the diameter of approximately 5 mm can be attributed to the trade-off between the degree of anisotropic characteristics and resistivity of the electrical adhesion (Figure 7(B)). Table II presents a comparison of the sensing performance of the single- and double-layer samples, with the results obtained experimentally. The findings highlight that the sensitivity decreases with stacking, and the position accuracy increases with the number of adhesive dots. These trends are consistent with those of the simulation study.

*3) Limitations and future work:* Notably, we implemented a tomographic tactile sensor based on resistive coupling using a porous media. However, the simulation results can be extended to other conductive materials, such as conductive polymer composites [38] and self-healing materials [39]. The sensitivity and maximum force are related to the elasticity of the detector. Therefore, the elasticity can be tuned for adjusting the sensitivity. However, the range of the elasticity is typically smaller than the that of the conductivity. The proposed framework for examining the relationship between the material properties and sensing performance can be used for other tomographic sensors, including those for multi-modal tomography [40], shape sensing [41], proximity imaging [42], and thermal imaging [43].

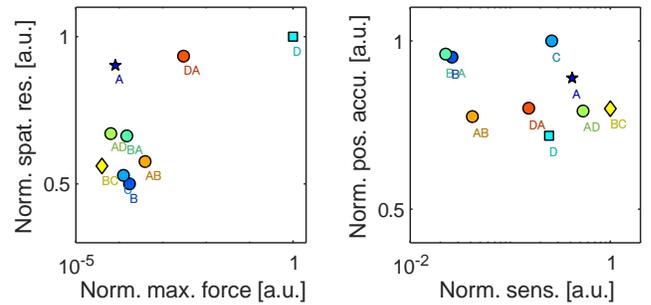

Fig. 6. Positioning plots obtained using experimental data (left) in the spatial resolution and maximum force space, (right) in the position accuracy and sensitivity space. These metrics are normalized.

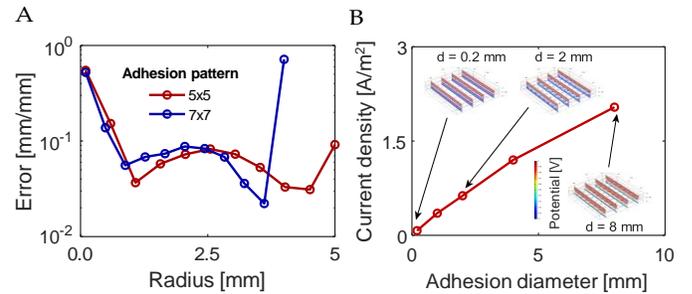

Fig. 7. Influence of the adhesion method. (A) Position error vs. dot diameter using 5 × 5 dots and 7 × 7 dots. The largest diameter of each dot pattern is equivalent to full surface adhesion. (B) Relationship between the adhesive diameter and current density at the adhesive surface. A DC voltage of 1 V between is applied between the top and bottom surfaces of the conductor with a conductivity of 1 S/m. The potential distributions are shown in the plot. The current density increases with the adhesive diameter. The adhesion characteristics were simulated using COMSOL Multiphysics 6.0.

## V. Demonstrations

To highlight the potential uses of the proposed sensor, we developed various types of detectors and demonstrated their applications in robotics, interfaces, and haptics. Both simulation and experimental results suggest that using a conductive porous material with a conductivity of approximately 0.2 S/m as a detection layer is generally effective for touch interactions of forces involving several Newtons. Therefore, the sensors for the demonstration were fabricated using Sample A, which exhibits a well-balanced performance as a detection layer.

A deformable sensor is valuable for a soft robots [44]. A deformable detector can be developed using a conductive porous media as the detection layer. Therefore, we designed a flexible, strip-shaped sensor, which could be attached to a soft gripper (Figure 8(A) and Movie S1).

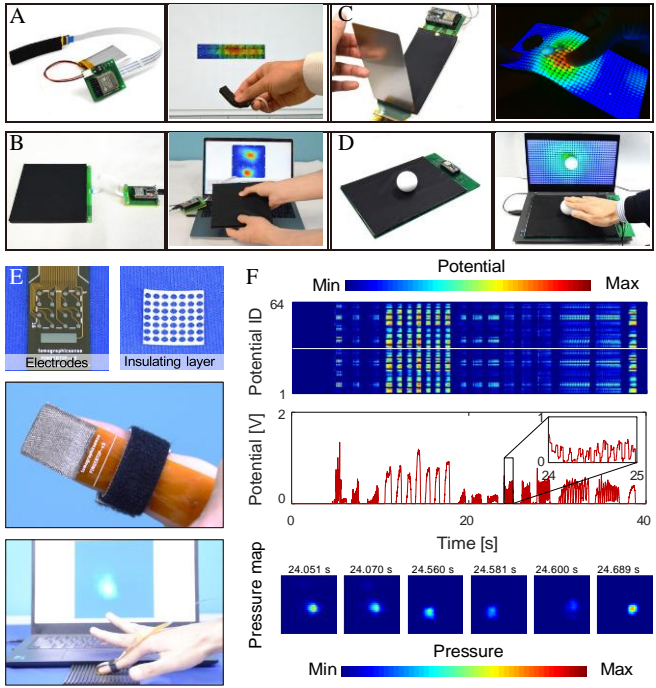

Fig. 8. Potential applications in robotics, interfaces, and haptics fields. (A) Flexible, strip-shaped sensor for a soft gripper. The detector has an area of 13 × 60 mm$^2$ and a thickness of 3 mm, incorporating 2 × 8 electrodes. (B) High resolution square sensor using the CNN-based reconstruction method. The detector has an area of 150 × 150 mm$^2$ and a thickness of 10 mm, incorporating 4 × 4 electrodes. (C) Flexible touch display using a 6 inch OLED display. The detector has an area of 140 × 70 mm$^2$ and a thickness of 3 mm, incorporating 8 × 4 electrodes. (D) Tangible user interface based on a ball and sensor pad. The detector has an area of 257 × 182 mm$^2$ and a thickness of 5 mm, incorporating 8 × 4 electrodes. (E) Wearable fingertip tactile sensor. The detector has an area of 10 × 10 mm$^2$ and a thickness of 0.5 mm, incorporating 4 × 4 electrodes. The potential vectors, extracted values at the 32 elements, and six frame captures of the pressure distribution in time-series are shown.

To enhance the spatial performance, a square sensor was developed using a CNN-based reconstruction method [32] with four grounding and four scanning techniques. The square sensor using the CNN-based reconstruction method exhibited superior spatial resolution (Figure 8(B) and Movie S2) compared to the sensor using a linear reconstruction method. Next, we demonstrated the use of a tactile sensor for realizing a flexible touch display. The tomographic tactile sensor was attached to the back side of a organic light-emitting diode (OLED) display. This system allowed users to interact with graphic contents displayed on the OLED display through direct touch (Figure 8(C) and Movie S3). As a novel tangible user interface [45], we implemented an input interface using a tactile sheet and a ball (Figure 8(D) and Movie S4). This system allowed users to operate a cursor by rolling and pressing the ball on the sensor sheet. Finally, a wearable fingertip tactile sensor was established (Figure 8(E) and Movie S5 and S6), which could be used for analyzing tactile stimuli at the fingertip with minimal constraints on the body. Figure 8(F) shows potential vectors, extracted values, and pressure distributions at certain time instants during free touch interaction with a periodically uneven rubber surface over 40 s. Because the sensor has a high spatial-temporal resolution i.e., <2 mm for two-point discrimination, and a sampling rate of 1000 kHz, it can be used for identifying material properties and touch actions. Specifically, the potential pattern at 24 – 25 s shows that the sensor can effectively capture vibration patterns (>10 Hz).

## VI. Conclusion

In this study, we obtained performance maps with the conductivity of a scalable detector for a contact resistance based tomographic tactile sensor by simulation, and confirmed actual performance using porous conductive materials with the selected conductivities. The performance maps revealed that using a material with a conductivity of approximately 0.2 S/m can serve as an effective detector for high-performance sensing in touch interactions involving a force range of several Newtons. Moreover, incorporating gradient conductivity in the cross-section of the detector and multi-layer conductive porous media with anisotropic conductive bonding can help expand the design flexibility for enhanced performance. Based on these findings, we demonstrated various high-performance tomographic tactile sensors for soft grippers, tangible input interfaces, flexible touch displays, and wearable electronics by using a conductive porous media. Our finding provides a major advance in design of soft tactile sensors, which can be used for robot control, analysis of human touch, input interface, haptic feedback, and wearable devices.

## Acknowledgment

The authors thanks Mitsuki Funato, Akira Kojima and Sotaro Hattori for the help during the research.